\documentclass[floatfix,10pt]{wlscirep}

\usepackage{float}

\newcommand{\p}{\partial}

\title{Exploring nonlinear topological states of matter with exciton-polaritons: Edge solitons in kagome lattice}

\author[1,2]{D. R. Gulevich}
\author[1,3]{D. Yudin}
\author[1,2]{D. V. Skryabin}
\author[1,3]{I. V. Iorsh}
\author[1,4]{I. A. Shelykh}
\affil[1]{ITMO University, St. Petersburg 197101, Russia}
\affil[2]{Department of Physics, University of Bath, Bath BA2 7AY, United Kingdom}
\affil[3] { Division of Physics and Applied Physics, Nanyang Technological University 637371, Singapore}
\affil[4]{Science Institute, University of Iceland, Dunhagi 3, IS-107, Reykjavik, Iceland}

\begin{abstract}
Matter in nontrivial topological phase possesses unique properties, such as support of unidirectional edge modes on its interface. It is the existence of such modes which is responsible for the wonderful properties of a topological insulator -- material which is insulating in the bulk but conducting on its surface, along with many of its recently proposed photonic and polaritonic analogues. We show that exciton-polariton fluid in a nontrivial topological phase in kagome lattice, supports nonlinear excitations in the form of solitons built up from wavepackets of topological edge modes -- topological edge solitons. Our theoretical and numerical results indicate the appearance of bright, dark and grey solitons dwelling in the vicinity of the boundary of a lattice strip. In a parabolic region of the dispersion the solitons can be described by envelope functions satisfying the nonlinear Schr\"odinger equation. 
Upon collision, multiple topological edge solitons emerge undistorted, which proves them to be true solitons as opposed to solitary waves for which such requirement is waived. Importantly, kagome lattice supports topological edge mode with zero group velocity unlike other types of truncated lattices. This gives a finer control over soliton velocity which can take both positive and negative values depending on the choice of forming it topological edge modes.
\end{abstract}
\begin{document}

\flushbottom
\maketitle

\thispagestyle{empty}

\section*{Introduction}

Various physical systems often demonstrate similarity in the underlying physical phenomena, which drives an idea of exploiting well controllable systems for mimicking properties of the less controllable and the less accessible ones.
A spectacular example is the similarity between electronic and photonic systems~\cite{Georgescu2014}. 
It is therefore not surprising that with the fast rise of topological insulators in the context of electronic systems~\cite{Kane-Mele,Bernevig,Konig}, topological ideas were 
also widely explored in photonic systems. Among the pioneering works are the  
study of chiral edge states in photonic crystals~\cite{Raghu-2008} where the Berry curvature for photonic bands was introduced by analogy of electronic systems, photonic analogues of Hall effect~\cite{Haldane2008a, Wang2008,Wang-2009} and topological insulators~\cite{Hafezi-2011,Fang-2012,Rechtsman-2013,Khanikaev2013}, as well as a solid number of both theoretical and experimental works demonstrating the effects of non-trivial topology in electromagnetic systems in the frequency range from radio to optics~\cite{Lu-2014}. 

Despite this success, however, optical circuits are not well suited for nonlinear effects to be directly incorporated while realization of the time-reversal symmetry breaking, a common ingredient of topological phases, remains challenging. 
In this respect, systems based on exciton-polaritons~\cite{Carusotto2013}, quasi-particles originating from the strong coupling of the quantum well-excitons and cavity photons in microcavities, are at advantage. 
Being hybrid light-matter excitations, their photonic properties allow an effective control with the use of optical potential profile, while their excitonic nature brings significant interactions and a strong nonlinear response~\cite{CedraMendez2010,Kim2013,Jacqmin2014,Baboux2016}. 
Moreover, due to the exciton spin, the time-reversal symmetry of an exciton-polariton system can be conveniently broken by application of the external magnetic field. This makes polaritonic systems attractive both from the point of view of applications in prospective polaritonic devices~\cite{Liew2011,Sanvitto2016} and as a unique laboratory to simulate topological properties of matter.

There had been several proposals for creating topologically nontrivial states of exciton-polaritons. In the last few years, emergence of the non-trivial topology and existence of the topologically protected edge states in polaritonic lattices of different geometry were addressed in a number of works~\cite{Karzig-PRX-2015, Bardyn-PRB-2015, Nalitov-Z, Yi-PRB-2016, Bardyn-PRB-2016, Janot-PRB-2016, Gulevich-kagome}.
Currently, the focus of attention in the study of the effects of non-trivial topology shifts towards systems with nonlinearity, where exciton-polaritons, due to their unique properties, play a special role. Among recent works where an interplay of nonlinear effects with nontrivial topology has been explored are studies of self-localized states~\cite{Lumer-PRL-2013, Ostrovskaya2013}, self-induced topological transitions~\cite{Hadad-PRB-2016}, topological Bogoliubov excitations~\cite{Bardyn-PRB-2016}, suppression of topological phases~\cite{Bleu-PRB-2016}, vortices in lattices~\cite{Kartashov-OL-2016}, spin-Meissner states in ring resonators~\cite{DGulevich-Meissner}, solitons in lattices~\cite{Ablowitz-PRA-2014, Leykam-PRL-2016, Kartashov-Optica-2016, CedraMendez2016} and dimer chains~\cite{Soln-PRL-2017}. In Ref.~\citeonline{Gulevich-kagome} it was observed that in a certain range of parameters kagome lattice possesses a highly nonlinear dispersion of topological edge state, with a well pronounced minimum and maximum inside the bulk gap. In the present paper we show that such peculiar dispersion leads to appearance of nonlinear edge excitations in the form of solitons. Such excitations turn out to be true solitons as opposed to solitary waves for which the requirement of restoring shape upon collision is waived~\cite{Rajaraman}. In contrast to solitons in the honeycomb lattice~\cite{Kartashov-Optica-2016} 
velocity of topological edge state solitons in kagome lattice can take values in a wide range from positive to negative 
depending on the choice of quasimomenta of the constituting topological edge modes.

\section*{Results}

\subsection*{Model of Polaritonic Kagome Lattice}

\begin{figure}[ht!]
	\begin{center}
\includegraphics[width=6.9in]{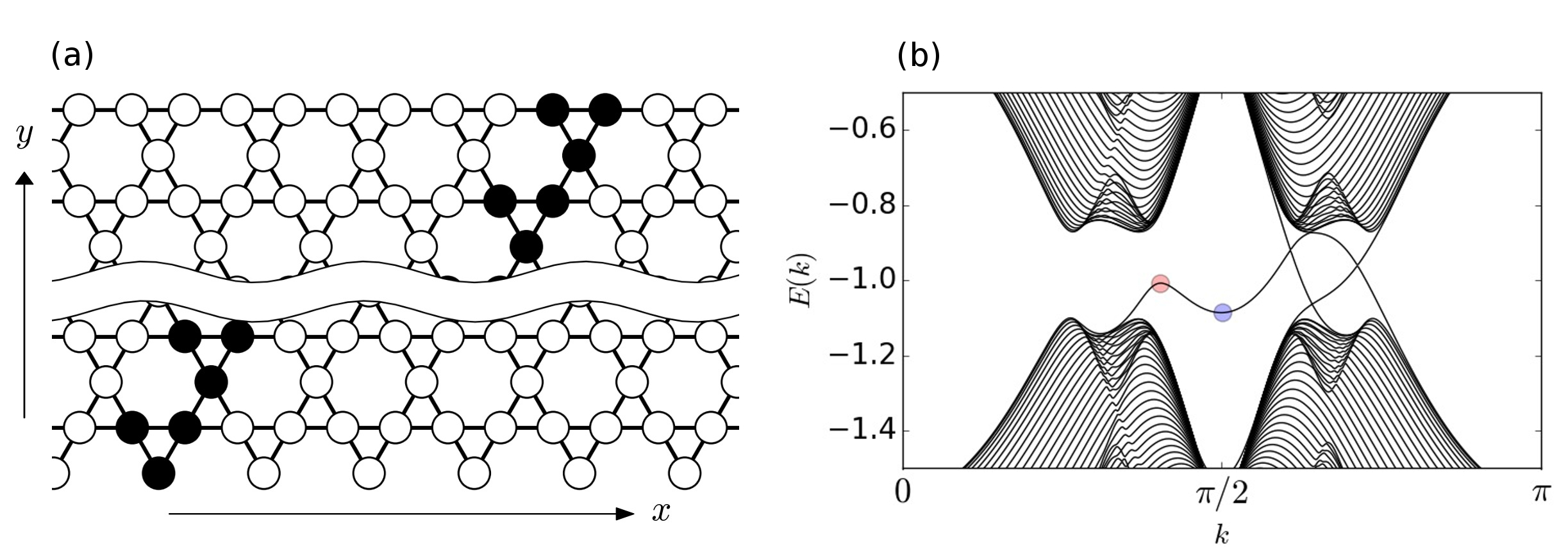}
\caption{\label{fig:strip-disp} 
(a) A strip of kagome lattice. The strip is infinite along $x$ but has a finite extent along the $y$ axis. The sites marked by black form a unit cell of the strip. The bottom boundary of the strip with a row of uncoupled sites supports the dispersion of topological edge mode with the group velocity changing sign within the bulk gap as shown in Figure b. 
(b) 
Band structure for exciton-polaritons in a strip of kagome lattice. 
Dispersion of edge states with reversed group velocity and a well defined minimum and maximum lying within the bulk gap appears for a certain range of values of the magnetic field and TE-TM splitting at the lower boundary made of sites without the nearest neighbor hopping (the bottom one in Figure a): in this example, $\Omega=0.3$ and $\delta J=0.15$ in normalized units. The red and blue dots mark the regions where nonlinear solutions in the form of bright and dark solitons are sought. The band structure beyond the edge of the first Brillouin zone at $\pi/2$ is shown for better display of the topological edge state dispersions.
}
\end{center}
\end{figure}

In the tight-binding approximation the Hamiltonian for polaritons confined to an array of coupled microcavity pillars arranged into a kagome lattice reads 
\begin{equation}
\hat{H}=
\Omega \sum_{i,\sigma}\sigma\,\hat{a}_{i,\sigma}^\dagger \hat{a}_{i,\sigma}
- \sum_{\langle ij\rangle,\sigma} \left( J \hat{a}_{i,\sigma}^\dagger \hat{a}_{j,\sigma} 
+ \delta J e^{2i\varphi_{ij}\bar\sigma} \hat{a}_{i,\sigma}^\dagger \hat{a}_{j,\bar\sigma} 
 + h.c.\right) 
+ \sum_{i,\sigma}\left( 
\frac{\alpha_1}{2}\, \hat{a}_{i,\sigma}^\dagger \hat{a}_{i,\sigma}
\hat{a}_{i,\sigma}^\dagger \hat{a}_{i,\sigma} 
+ \alpha_2\, \hat{a}_{i,\sigma}^\dagger \hat{a}_{i,\sigma}
\hat{a}_{i,\bar\sigma}^\dagger \hat{a}_{i,\bar\sigma} \right).
\label{tb}
\end{equation}
Equation~\eqref{tb} is a generalization of the linear model studied in Ref.~\citeonline{Gulevich-kagome} to account for polariton-polariton interactions. Here, operators $\hat{a}_{i,\sigma}^\dagger$ ($\hat{a}_{i,\sigma}$) create (annihilate) exciton-polariton of circular polarization $\sigma=\pm$ at site $i$ of the kagome lattice, the summation $\langle ij\rangle$ is over nearest neighbors (NN), angles $\varphi_{ij}$ specify directions of vectors connecting the neighboring sites. The first term in~\eqref{tb} describes the Zeeman energy splitting ($2\Omega$) of the circular polarized components induced by external magnetic field,
the second term describes the NN hopping with conservation (term with $J$) and inversion (term with $\delta J$) of circular polarization, and the last term describes the on-site polariton-polariton interactions with effective constants $\alpha_1$ and $\alpha_2$ defined for the given pillar mode. 
The NN coupling with inversion of polarization arises due to 
the linear polarization modes in neighboring pillars experiencing different tunnel barriers in presence of TE-TM splitting of the photonic modes~\cite{Vasc-APL-2011,Nalitov-Z,Nalitov-PRL-2015,Sala-PRX-2015,Gulevich-kagome}.
In what follows, we will use normalized units where $J$ is a unit energy and the interpillar distance is the unit length (size of the kagome lattice unit cell then equals 2 in these units, see Fig.~\ref{fig:strip-disp}a).

In the linear regime when the interactions in~\eqref{tb} can be neglected, the band structure of polaritonic kagome lattice was calculated in Ref.~\citeonline{Gulevich-kagome}. 
While the effect of $\Omega$ on the band structure is to shift energy dispersions of the two circular polarization by the amount of Zeeman splitting, the non-zero $\delta J$ results in 
coupling of the two circular polarizations and an anticrossing of the corresponding dispersion curves. Presence of both non-zero TE-TM splitting $\delta J$ and magnetic field $\Omega$ leads to a band inversion associated with a topological phase transition and opening of a gap which separates the Bloch bands into two bundles, each possessing a nontrivial topology~\cite{Gulevich-kagome}. The bulk-boundary correspondence from the theory of topological insulators (see, e.g. reviews~\cite{Hasan-RMP-2010, Fruchart}) suggests that on the interface between a topologically non-trivial phase and a topologically trivial one (or, vacuum), there must exist topological edge modes (TEM).

While the number of TEM is a topological invariant and, therefore, is independent of the type of a boundary, the very dispersion of TEM does depend on how the lattice is cut at its interface. The boundary with a row of pillars uncoupled between each other (see, the bottom edge of the strip in Fig.~\ref{fig:strip-disp}a) is more prone to localization and thus generally exhibits flatter dispersions of TEM compared to those where the sites are directly connected via NN coupling (e.g. the top edge on Fig.~\ref{fig:strip-disp}a). Furthermore, the sign of the group velocity of edge states propagating along the ``uncoupled'' boundary (the bottom one on Fig.~\ref{fig:strip-disp}a) can be reversed, see Fig.~\ref{fig:strip-disp}b. Such highly nonlinear dispersion turns out to be favorable for the existence of nonlinear excitations of TEM in the form of solitons. In the following sections we will show that this is indeed the case and present our analytic and numerical results.

\subsection*{Theory of nonlinear topological edge excitations in a 2D lattice}

To study nonlinear topological edge excitations we employ the mean field approximation. We consider a strip of kagome lattice with $M$ sites in its unit cell, see Fig.~\ref{fig:strip-disp}a, and introduce vectors $\pmb{\psi}_m^{\sigma}=(\psi_{m,1}^{\sigma},...,\psi_{m,M}^{\sigma})^T$ composed of the spinor components $\sigma=\pm$ at each site of the $m$th unit cell, where index $m$ enumerates the unit cells in the $x$ direction (see Fig.~\ref{fig:strip-disp}a). 
Then, the time evolution of the $m$th vector is described by
the coupled system of equations
\begin{equation}
i\p_t\pmb{\psi}_m^{\sigma}=\sum_{\tau}\left(\hat H_{-1}^{\sigma,\tau}\pmb{\psi}_{m-1}^{\tau}+\hat H_0^{\sigma,\tau}\pmb{\psi}_m^{\tau}+\hat H_{+1}^{\sigma,\tau}\pmb{\psi}_{m+1}^{\tau}\right)+ \hat{N}^{\sigma}(\pmb{\psi}_m^{+},\pmb{\psi}_m^{-}) \pmb{\psi}_m^{\sigma},
\label{td-problem}
\end{equation} 
where $\hat{N}^{\sigma}(\pmb{\psi}_m^{+},\pmb{\psi}_m^{-})$ is the diagonal matrix with elements
$$\left[\hat{N}^{\sigma}(\pmb{\psi}_m^{+},\pmb{\psi}_m^{-})\right]_{ij} = \delta_{i,j}\left(|\psi_{m,j}^{\sigma}|^2 + \alpha |\psi_{m,j}^{\bar\sigma}|^2\right),$$
parameter $\alpha\equiv \alpha_2/\alpha_1$, and 
$\pmb{\psi}_m^{\sigma}$ are normalized to $\sum_{\sigma,m}\langle\pmb{\psi}_m^{\sigma},\pmb{\psi}_m^{\sigma}\rangle \equiv \sum_{\sigma,m,j}[\pmb{\psi}_m^{\sigma*}]_j [\pmb{\psi}_m^{\sigma}]_j=N/\alpha_1$ 
with $N$ being the total number of polaritons in the system.
Matrices $\hat H_{j}^{\sigma,\tau}$, satisfy $(\hat H_{j}^{\sigma,\tau})^\dagger=\hat H_{-j}^{\tau,\sigma}$
for all $\sigma,\tau=\pm$, $j=0,\pm1$, which ensures hermiticity of the Hamiltonian.
In absence of interactions the stationary states of~\eqref{td-problem} are Bloch waves $e^{-i\mu t+ikLm}\,\pmb{u}_{k}^{\sigma}$ where $L$ is the spatial extent of the unit cell in $x$ direction and $\pmb{u}_{k}^{\sigma}$ are 
solutions to the eigenvalue problem
\begin{equation}
\sum_{\tau}\left(e^{-ikL} \hat H_{-1}^{\sigma,\tau}+\hat H_0^{\sigma,\tau}+e^{ikL} \hat H_{+1}^{\sigma,\tau}\right) \pmb{u}_{k,n}^{\tau}=\mu_{k,n}\,\pmb{u}_{k,n}^{\sigma}
\quad\text{for $\sigma=\pm$,}
\label{operator}
\end{equation}
where $n$ enumerates the energy bands. Because the operator in the left hand side of~\eqref{operator} is self-adjoint, $\pmb{u}_{k,n}^{\sigma}$ forms an orthonormal set, 
$$
\sum_\sigma\langle\pmb{u}_{k,n}^{\sigma},\pmb{u}_{k,m}^{\sigma}\rangle 
=\sum_{\sigma,j}[\pmb{u}_{k,n}^{\sigma*}]_j[\pmb{u}_{k,m}^{\sigma}]_j=\delta_{n,m}.
$$
Let $n=n_e$ is the band index corresponding to one of the TEM dispersions as in Fig.~\ref{fig:strip-disp}b.
Then, solution to the full problem~\eqref{td-problem} can be sought in the form of a wavepacket centered around $k_e$,
\begin{equation}
\pmb{\psi}^{\sigma}_m(k_e,t)
=\sum_n \int_{-\pi/L}^{\pi/L} A_n(\kappa,t)\pmb{u}^{\sigma}_{k_e+\kappa,n} e^{i(k_e+\kappa)Lm} d\kappa
\approx \int_{-\pi/L}^{\pi/L} A(\kappa,t)\pmb{u}^{\sigma}_{k_e+\kappa,n_e} e^{i(k_e+\kappa)Lm} d\kappa,
\label{psi-anzatz}
\end{equation}
where we assumed that a single TEM with $n=n_e$ and amplitude $A(\kappa,t)$ dominates other modes.
For the amplitude $A(\kappa,t)$ 
to be uniquely defined, one needs to fix a gauge of the basis vectors $\pmb{u}^{\sigma}_{k,n_e}$. It is always possible to choose a gauge such that the equation
\begin{equation}
\sum_{\sigma}\langle\pmb{u}^{\sigma}_{k,n_e},\frac{\p }{\p k}\pmb{u}^{\sigma}_{k,n_e}\rangle=0
\label{gauge}
\end{equation}
is satisfied at least within an 
open interval of $k$ smaller than the Brillouin zone. While the real part of~\eqref{gauge} is guaranteed by the normalization, the imaginary part of~\eqref{gauge} can be forced by the $U(1)$ rotation of eigenvector phases. 
Substituting~\eqref{psi-anzatz} to~\eqref{td-problem} and forming a scalar product with $\pmb{u}^{\sigma}_{k_e,n_e}$ we get
\begin{equation}
\int_{-\pi/L}^{\pi/L}\sum_\sigma
\left[ \left(i\frac{\p A(\kappa,t)}{\p t}-\mu_{k_e+\kappa,n_e}A(\kappa,t)\right)\langle\pmb{u}^{\sigma}_{k_e,n_e},\pmb{u}^{\sigma}_{k_e+\kappa,n_e}\rangle
- A(\kappa,t)\langle\pmb{u}^{\sigma}_{k_e,n_e},\hat{N}^{\sigma}(\pmb{\psi}_m^{+},\pmb{\psi}_m^{-})\pmb{u}^{\sigma}_{k_e+\kappa,n_e}\rangle 
\right] e^{i\kappa Lm} d\kappa = 0
\label{nls-0}
\end{equation}
for all $m$. 
By differentiating~\eqref{gauge} we obtain
\begin{equation}
\sum_{\sigma}\langle\frac{\p }{\p k}\pmb{u}^{\sigma}_{k,n_e},\frac{\p }{\p k}\pmb{u}^{\sigma}_{k,n_e}\rangle+
\sum_{\sigma}\langle\pmb{u}^{\sigma}_{k,n_e},\frac{\p^2 }{\p^2 k}\pmb{u}^{\sigma}_{k,n_e}\rangle=0.
\label{p2k}
\end{equation}
The gauge condition~\eqref{gauge} 
implies that the derivative $\p\pmb{u}^{\sigma}_{k,n_e}/\p k$ comprises excitations of all other bands but $n_e$ (in general, the bulk bands and the other TEM branches). In particular, at $k=k_e$,
\begin{equation}
\frac{\p }{\p k}\pmb{u}^{\sigma}_{k,n_e}\Big|_{k=k_e} = \sum_{n\neq n_e} c_n \pmb{u}^{\sigma}_{k_e,n}.
\label{cn}
\end{equation}
Expanding $\pmb{u}^{\sigma}_{k_e+\kappa,n_e}$ into the Maclaurin series in $\kappa$
and using~\eqref{p2k} and~\eqref{cn}, we get
\begin{equation}
\sum_{\sigma}\langle\pmb{u}^{\sigma}_{k_e,n_e},\pmb{u}^{\sigma}_{k_e+\kappa,n_e}\rangle\approx 1-\frac12 \kappa^2\sum_{n\neq n_e}|c_n|^2.
\label{cond}
\end{equation}
Assuming the excitation of bands $n\neq n_e$ is weak, which is guaranteed by the large energy separation of the band $n_e$ from the rest of the bands, we can neglect
the contribution of the bulk bands in~\eqref{nls-0}.
Decomposing $\mu_{k_e+\kappa,n_e}$ into the Maclaurin series up to the 2nd order in $\kappa$ and integrating, the Eq.~\eqref{nls-0} is reduced to
\begin{equation}
i\frac{\p \tilde{A}}{\p t} =
\sum_{n=0}^{\infty} \frac{(-i)^n}{n!} \mu_{k_e}^{(n)}\frac{\partial^n \tilde{A}}{\partial x^n}\Big|_{x=Lm}
+ g |\tilde{A}|^2 \tilde{A}\quad\text{for all $x=Lm$},
\label{nls-2}
\end{equation}
where 
$$
\tilde{A}(x,t)\equiv \int_{-\pi/L}^{\pi/L} A(\kappa,t) e^{i\kappa x}  d\kappa
$$
is a function of time and continuous variable $x$, and 
$$
g\equiv \sum_\sigma\langle\pmb{u}^{\sigma}_{k_e,n_e},\hat{N}^{\sigma}(\pmb{u}_{k_e,n_e}^+,\pmb{u}_{k_e,n_e}^{-})\pmb{u}^{\sigma}_{k_e,n_e}\rangle
$$
is the effective nonlinearity parameter. For $\alpha \ge -1$, which is the usual case for polariton interactions~\cite{Vladimirova}, $g$ can be proved to take non-negative values only, $g\ge 0$, irrespective of~$\pmb{u}_{k_e,n_e}^\sigma$, thus describing the defocusing nonlinearity.
Finally, 
\begin{equation}
\pmb{\psi}^{\sigma}_m(k_e,t) \approx  \tilde{A}(Lm,t)\, e^{i k_e Lm}\, \pmb{u}^{\sigma}_{k_e,n_e}.
\label{psi-final}
\end{equation}

Using~\eqref{cond}, we can analyze the validity of the applied approximation. 
Coefficients $c_n$ can be calculated from the standard perturbation theory~\cite{LL-III}, which gives
\begin{equation}
\quad
c_n = \frac{1}{\mu_{k_e,n_e}-\mu_{k_e,n}}\,
\sum_{\sigma,\tau} \langle \pmb{u}^{\sigma}_{k_e,n}, \hat{V}^{\sigma,\tau}
\pmb{u}^{\tau}_{k_e,n_e}
\rangle,
\quad
\hat{V}^{\sigma,\tau} = iL \left(e^{ik_eL} \hat H_{+1}^{\sigma,\tau} - e^{-ik_eL} \hat H_{-1}^{\sigma,\tau}\right).
\label{cn-pert}
\end{equation}
Thus, we can estimate
\begin{multline}
\sum_{n\neq n_e}|c_n|^2 \le \frac{1}{\Delta\mu^2} 
\sum_{n\neq n_e} 
\left|\sum_{\sigma,\tau} \langle \pmb{u}^{\sigma}_{k_e,n}, \hat{V}^{\sigma,\tau}\pmb{u}^{\tau}_{k_e,n_e} \rangle\right|^2
=
\frac{1}{\Delta\mu^2} \left(
\sum_{\sigma,\tau,\tau'} 
\langle \hat{V}^{\sigma,\tau}\pmb{u}^{\tau}_{k_e,n_e} , \hat{V}^{\sigma,\tau'}\pmb{u}^{\tau'}_{k_e,n_e}\rangle
- \left| \sum_{\sigma,\tau} \langle \pmb{u}^{\sigma}_{k_e,n_e}, \hat{V}^{\sigma,\tau}\pmb{u}^{\tau}_{k_e,n_e} \rangle \right|^2
\right) 
\\
\le \frac{1}{\Delta\mu^2}
\sum_{\sigma,\tau,\tau'} 
\langle \hat{V}^{\sigma,\tau}\pmb{u}^{\tau}_{k_e,n_e} , \hat{V}^{\sigma,\tau'}\pmb{u}^{\tau'}_{k_e,n_e}\rangle,
\label{estimate}
\end{multline}
where we introduced the energy separation from the bulk modes $\Delta\mu\equiv {\rm min}_{n\neq n_e} |\mu-\mu_n|$
and used completeness of the basis set. 
The last term in~\eqref{estimate} can be estimated using the exact form~\eqref{cn-pert} for the operator $\hat{V}^{\sigma,\tau}$ keeping only the hopping terms with conserved circular polarization in the original Hamiltonian~\eqref{tb}.
This yields an estimate $\sim L^2/\Delta\mu^2$.
Defining the average
$$
\langle f(\kappa) \rangle_{A} \equiv \left( \int_{-\pi/L}^{\pi/L} |A(\kappa,t)| d\kappa\right)^{-1} \int_{-\pi/L}^{\pi/L} |A(\kappa,t)| f(\kappa) d\kappa
$$
and using~\eqref{cond},~\eqref{estimate} and $L=2$ we can obtain the criteria for validity of~\eqref{nls-2} and~\eqref{psi-final},
\begin{equation}
2\langle \kappa^2 \rangle_{A} \ll \Delta\mu^2,
\label{cond-final}
\end{equation}
which is a restriction on the extent of wavepacket in k-space for a given energy separation from the bulk modes.
Note, that an obvious requirement of the independence of~\eqref{psi-final} of the choice of the kagome strip unit cell (note, that the choice of the unit cell in Fig.~\ref{fig:strip-disp} is not unique) requires either a weak dependence of $\tilde{A}(x,t)$ on $x$ (that is, a small $\langle \kappa^2 \rangle_{A}$), or, a strong localization of the edge mode 
near the boundary within the first few rows (ensured by large $\Delta\mu^2$).

\subsection*{Topological Edge Solitons}
If the higher order terms in~\eqref{nls-2} can be neglected, the equation can be 
further reduced to the standard nonlinear Schr\"{o}dinger (NLS) equation. Disregarding the terms with $n>2$, the solution to~\eqref{nls-2} can be written as
\begin{equation}
\tilde{A}(x,t) = \exp\left\{-i\mu_{k_e} t+i\frac{(v-\mu'_{k_e})}{\mu''_{k_e}}\left[x-\frac12(v+\mu'_{k_e}) t\right]\right\} a(x-vt,t),
\label{Aa}
\end{equation}
where $v$ is an arbitrary real parameter, 
and $a(x,t)$ satisfies the NLS equation 
\begin{equation}
i\frac{\p a}{\p t} + \frac{1}{2} \mu''_{k_e} \frac{\partial^2 a}{\partial x^2} - g |a|^2 a=0.
\label{nls-3}
\end{equation}
Note that~\eqref{Aa} takes a form of Galilean transformation on the solution $a(x,t)$ of the NLS equation, thus introducing dynamics of the NLS solution $a(x,t)$ with velocity $v$.
For $\mu''_{k_e}<0$ Eq.~\eqref{nls-3} allows solutions in the form of bright soliton (see, e.g. Ref.~\citeonline{Ablowitz}),
\begin{equation}
a_{\rm bright}(x,t) = \frac{\eta \,e^{-i\frac12 \eta^2 g t}}{\cosh\left({\eta\sqrt{\frac{g}{|\mu''_{k_e}|}}\, x}\right)},
\label{bright}
\end{equation}
where $\eta$ is an arbitrary real parameter.
For $\mu''_{k_e}>0$, there arise dark,
\begin{equation}
a_{\rm dark}(x,t) = \eta e^{-i \eta^2 g t} \tanh\left[\eta \sqrt{\frac{g}{\mu''_{k_e}}}\; x \right],
\label{dark}
\end{equation}
and, more general, grey solitons
\begin{equation}
a_{\rm grey}(x,t) = \eta e^{-i \eta^2 g t}\left\{\cos\theta+i\sin\theta\;\tanh\left[\left(\sqrt{\frac{g}{\mu''_{k_e}}}\;x+\eta\cos\theta\; g t\right)\eta \sin\theta\right] \right\},
\label{grey}
\end{equation}
where $\eta$ and $\theta$ are real parameters~\cite{Ablowitz}. 
Note, that the grey soliton envelope~\eqref{grey} introduces a phase shift $2|\theta|{\rm sgn}(\eta)$ between the amplitudes at limiting points $x\to\pm\infty$.
As seen from~\eqref{Aa}, 
the parameter $v$ plays role of the soliton velocity for the bright and dark solitons of the Eq.~\eqref{nls-2}, while the velocity of grey soliton, upon the transformation~\eqref{Aa} is
\begin{equation}
v_{\rm grey}=v-\eta\sqrt{g\mu''_{k_e}}\,\cos\theta.
\label{vgrey}
\end{equation}

\begin{figure}[!ht]
	\begin{center}
	\includegraphics[width=6in]{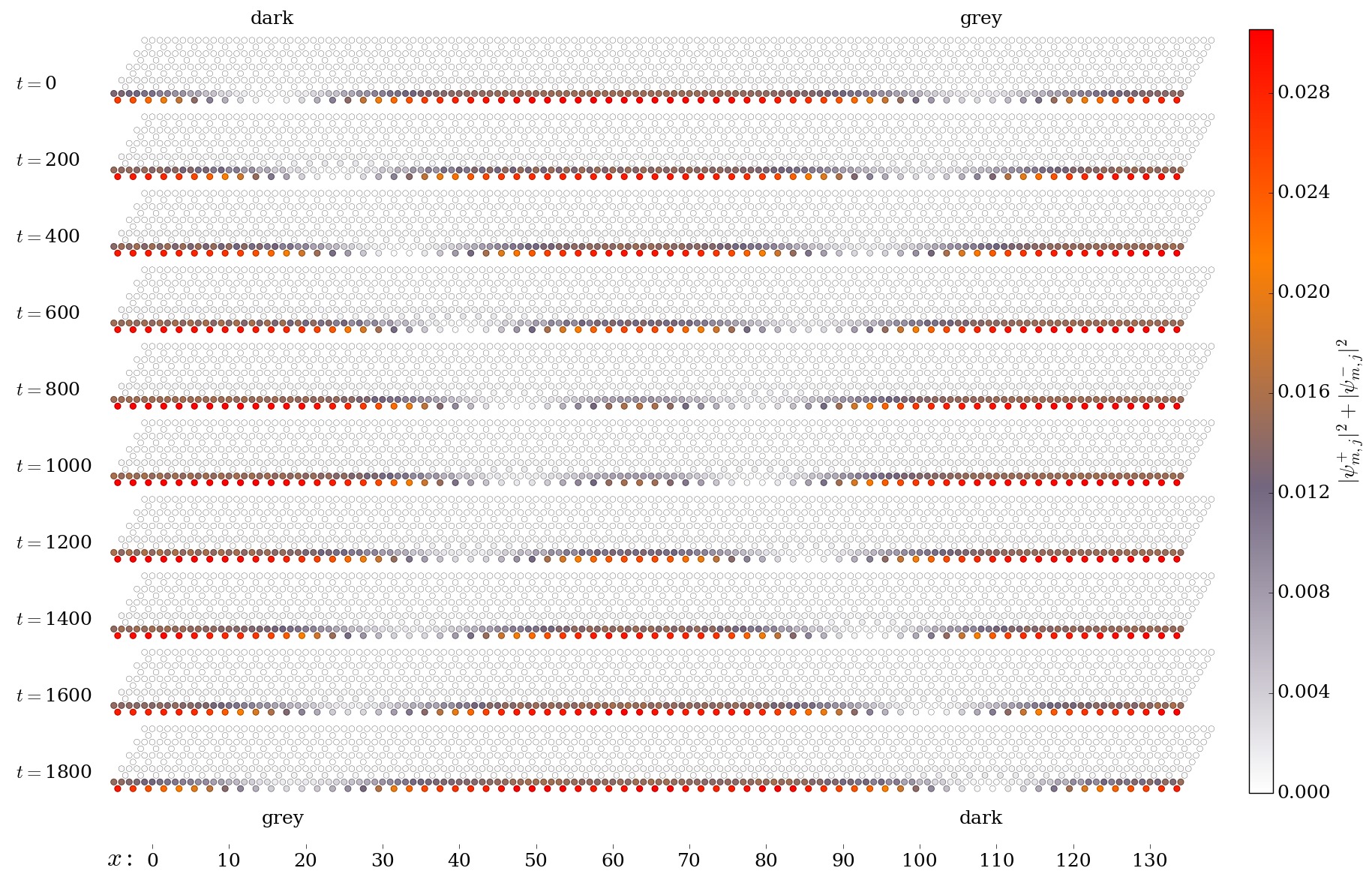}
\caption{\label{fig:num} 
(Color online) Simulation of collision of dark and grey solitons excited on the TEM propagating along the lower boundary of the kagome lattice strip in Fig.~\ref{fig:strip-disp}a. The TEM is excited at quasimomentum $k_e=1.59$, $E(k_e)=-1.09$ at $\delta J=0.15$, $\Omega=0.3$, in the vicinity of the blue spot on Fig.~\ref{fig:strip-disp}b. At $t=0$ initial conditions in the form of dark~\eqref{dark} (on the left) and grey~\eqref{grey} soliton (on the right) with $\eta=0.25$ have been taken upon applying the transformation~\eqref{Aa} with $v=0.042$. The grey soliton parameter $\theta$ was taken to be $0.4\pi$. At $t>0$ the dark and grey move with constant velocities until the scattering act occurs at $t\approx 900$, following which the dark and grey solitons exchange sides.
}
\end{center}
\end{figure}

\subsection*{Numerical results}

We have performed simulations of the dynamics of topological edge solitons by solving the mean field model~\eqref{td-problem} numerically. 
Our numerical calculations confirm the appearance of bright solitons of the type~\eqref{bright} whose velocities depend on the choice of the soliton momentum~$k_e$ with respect to the peak of the TEM dispersion on Fig.~\ref{fig:strip-disp}b. Due to a small energy separation from the bulk modes near the maximum of the TEM dispersion,
our analytical approach described above
works well only for very extended soliton profiles localized on a few dozens of unit cells.
On the other hand, a well separated from the bulk modes, nearly perfect parabolic, dispersion occurs near the edge of the first Brillouin zone at $k_e\approx \pi/2$, see Fig.~\ref{fig:strip-disp}b.
Owing to the positive second derivative $\mu''_{k_e,n_e}>0$,
it is a region where dark and grey solitons in the form~\eqref{dark} and~\eqref{grey} are to be observed. 

In our numerical simulations of the dynamics of dark and grey solitons, presented in Fig.~\ref{fig:num}, the TEM is excited at $k_e=1.59$, $E(k_e)=-1.09$, $\delta J=0.15$, $\Omega=0.3$, in the vicinity of the blue spot on Fig.~\ref{fig:strip-disp}b. The TEM dispersion $\mu_{k,n_e}$ and the eigenstate $\pmb{u}_{k_e,n_e}^{\sigma}$ were found by solving numerically the eigenvalue problem~\eqref{operator}. At the chosen value of $k_e$, the second derivative of the dispersion is $\mu''_{k_n,n_e}\approx 3.06$ and the nonlinearity parameter $g\approx 0.33$ where we assumed $\alpha=-0.05$ which is typical case for polariton-polariton interactions.
At the initial moment the TEM is tempered by the dark and grey soliton envelopes according to the Eqs.~\eqref{psi-final},~\eqref{Aa} with soliton profiles given by~\eqref{dark},~\eqref{grey} with $\eta=0.25$, and excited at a distance from each other. The Galilean transformation~\eqref{Aa} has been applied with parameter $v=0.042$, forcing the dark soliton to move to the right with velocity $v_{dark}=v$ according to~\eqref{Aa}. The parameter $\theta$ for the grey soliton was taken to be $0.4\pi$ which results in a grey soliton moving to the left with velocity $v_{grey}\approx -0.036$ upon transformation~\eqref{Aa}, according to~\eqref{vgrey}. As seen from our numerical calculations on Fig.~\ref{fig:num}, at $t>0$ the dark and grey solitons move with constant velocities as set by~\eqref{Aa},~\eqref{dark} and~\eqref{grey} while keeping their shapes during propagation. Upon reaching $t\approx 900$, scattering act occurs upon which dark and grey solitons change sides, see Fig.~\ref{fig:num}. It it interesting to note, that no actual passing of solitons {\it through} each other occurs. Instead, the interacting solitons stop at a distance from each other while exchanging their phase shifts. Upon the interaction, the emerged dark soliton propagates to the right, while the emerged grey soliton keeps propagating to the left.

\section*{Discussion}

Because the studied nonlinear topological edge excitations are formed from the TEM with energies inside the bulk gap, the arising solitons are robust against interaction with the bulk modes.
Note, that bright topological edge solitons arising here should be distinguished from the gap solitons. Being formed from topological edge modes, bright topological edge solitons can only propagate along the boundaries of the lattice, in contrast to gap solitons, which propagate inside the bulk. 

We predict that in the setting presented here it will be easier to detect experimentally dark and grey solitons rather than the bright ones. Indeed, as seen from Fig.~\ref{fig:strip-disp}b, at a fixed $k_e$ the separation from the bulk modes is much larger at the minimum of the TEM dispersion (at $k_e\approx \pi/2$) as opposed to the maximum of the TEM dispersion (at $k_e\approx 1.3$). Therefore, the condition~\eqref{cond-final} is easier to satisfy for observation of dark and grey solitons. As an example, at $k\approx \pi/2$ 
the separation between the $n_e-1$ and $n_e+1$ bands amounts $1.2$ in normalized units. Note, however, that the criterion~\eqref{cond-final} does not prohibit the existence of solitons in the regime when it is violated, rather, indicates that the nonlinear excitations can not be described in a simple mathematical form~\eqref{psi-final} with envelopes satisfying the NLS-type equation~\eqref{nls-2}.

In a possible experiment for observation of dark and grey edge solitons, the TEM should be first  excited by irradiating the lattice with laser
(see, e.g., recent experiment~\cite{Milicevic} where edge states in polaritonic honeycomb lattice were observed).
Once the TEM is excited, a distortion of the pumped radiation should be introduced whose magnitude and spatial profile will define the size and number of the emerging dark and grey solitons. 
The velocity of propagation of solitons can be controlled by the angle of incidence and frequency of the pumped radiation. In order to excite topologically protected edge states one needs to open energy gap in the band structure. 
Value of the coupling strength $J$ for experimentally fabricated microcavity pillars typically lies in the range $0.1$ to about $1\;\rm meV$~\cite{Galbiati,Abbarchi,Jacqmin2014,Milicevic} and can even reach 2.5 meV in open cavity systems~\cite{Duff-APL}. For $J=700\;\rm \mu eV $ realization of the dispersion as in Figure 1 requires $\Omega\approx 200\;\mu eV$ and $\delta J\approx 100\;\mu eV$, which is within the experimental reach. Moreover, opening a large gap is not necessary if topological dark and grey edge solitons are to be observed. Indeed, our numerical calculations show that propagation of dark and grey edge solitons is possible even when the minimum of the topological edge state dispersion falls below the energy of the bulk bands, e.g. at $\Omega=\delta J=70\;\mu eV$, $J=700\rm\;\mu eV$, and 
even further, until the gap collapses at $\Omega=\delta J=0$. 
In this regime, however, the edge modes on which dark and grey solitons reside are no longer topologically protected and can be prone to polariton parametric scattering processes~\cite{Kuwata-PRL-1997,Savvidis-PRL-2000, Ciuti-PRB-2001, Dasbach-PRB-2001, Tartakovskii-PRB-2002}.


\begin{thebibliography}{10}
	\expandafter\ifx\csname url\endcsname\relax
	\def\url#1{\texttt{#1}}\fi
	\expandafter\ifx\csname urlprefix\endcsname\relax\def\urlprefix{URL }\fi
	\expandafter\ifx\csname doiprefix\endcsname\relax\def\doiprefix{DOI }\fi
	\providecommand{\bibinfo}[2]{#2}
	\providecommand{\eprint}[2][]{\url{#2}}
	
	\bibitem{Georgescu2014}
	\bibinfo{author}{Georgescu, I.}, \bibinfo{author}{Ashhab, S.} \&
	\bibinfo{author}{Nori, F.}
	\newblock \bibinfo{title}{Quantum simulation}.
	\newblock \emph{\bibinfo{journal}{Reviews of Modern Physics}}
	\textbf{\bibinfo{volume}{86}}, \bibinfo{pages}{153} (\bibinfo{year}{2014}).
	
	\bibitem{Kane-Mele}
	\bibinfo{author}{Kane, C.~L.} \& \bibinfo{author}{Mele, E.~J.}
	\newblock \bibinfo{title}{Quantum spin hall effect in graphene}.
	\newblock \emph{\bibinfo{journal}{Phys. Rev. Lett.}}
	\textbf{\bibinfo{volume}{95}}, \bibinfo{pages}{226801}
	(\bibinfo{year}{2005}).
	
	\bibitem{Bernevig}
	\bibinfo{author}{Bernevig, B.~A.}, \bibinfo{author}{Hughes, T.~L.} \&
	\bibinfo{author}{Zhang, S.~C.}
	\newblock \bibinfo{title}{Quantum spin hall effect and topological phase
		transition in hgte quantum wells}.
	\newblock \emph{\bibinfo{journal}{Science}} \textbf{\bibinfo{volume}{314}},
	\bibinfo{pages}{1757} (\bibinfo{year}{2006}).
	
	\bibitem{Konig}
	\bibinfo{author}{K\"{o}nig, M.} \emph{et~al.}
	\newblock \bibinfo{title}{Quantum spin hall insulator state in hgte quantum
		wells}.
	\newblock \emph{\bibinfo{journal}{Science}} \textbf{\bibinfo{volume}{318}},
	\bibinfo{pages}{766} (\bibinfo{year}{2007}).
	
	\bibitem{Raghu-2008}
	\bibinfo{author}{Raghu, S.} \& \bibinfo{author}{Haldane, F. D.~M.}
	\newblock \bibinfo{title}{Analogs of quantum-hall-effect edge states in
		photonic crystals}.
	\newblock \emph{\bibinfo{journal}{Phys. Rev. A}} \textbf{\bibinfo{volume}{78}},
	\bibinfo{pages}{033834} (\bibinfo{year}{2008}).
	
	\bibitem{Haldane2008a}
	\bibinfo{author}{Haldane, F.} \& \bibinfo{author}{Raghu, S.}
	\newblock \bibinfo{title}{Possible realization of directional optical
		waveguides in photonic crystals with broken time-reversal symmetry}.
	\newblock \emph{\bibinfo{journal}{Phys. Rev. Lett.}}
	\textbf{\bibinfo{volume}{100}}, \bibinfo{pages}{013904}
	(\bibinfo{year}{2008}).
	
	\bibitem{Wang2008}
	\bibinfo{author}{Wang, Z.}, \bibinfo{author}{Chong, Y.},
	\bibinfo{author}{Joannopoulos, J.~D.} \& \bibinfo{author}{Solja{\v{c}}i{\'c},
		M.}
	\newblock \bibinfo{title}{Reflection-free one-way edge modes in a gyromagnetic
		photonic crystal}.
	\newblock \emph{\bibinfo{journal}{Phys. Rev. Lett.}}
	\textbf{\bibinfo{volume}{100}}, \bibinfo{pages}{013905}
	(\bibinfo{year}{2008}).
	
	\bibitem{Wang-2009}
	\bibinfo{author}{Wang, Z.}, \bibinfo{author}{Chong, Y.},
	\bibinfo{author}{Joannopoulos, F.~D.} \& \bibinfo{author}{Solja\v{c}i\'{c},
		M.}
	\newblock \bibinfo{title}{Observation of unidirectional backscattering-immune
		topological electromagnetic states}.
	\newblock \emph{\bibinfo{journal}{Nature}} \textbf{\bibinfo{volume}{461}},
	\bibinfo{pages}{772--775} (\bibinfo{year}{2009}).
	
	\bibitem{Hafezi-2011}
	\bibinfo{author}{Hafezi, M.}, \bibinfo{author}{Demler, E.~A.},
	\bibinfo{author}{Lukin, M.~D.} \& \bibinfo{author}{Taylor, J.~M.}
	\newblock \bibinfo{title}{Robust optical delay lines with topological
		protection}.
	\newblock \emph{\bibinfo{journal}{Nature Physics}}
	\textbf{\bibinfo{volume}{7}}, \bibinfo{pages}{907--912}
	(\bibinfo{year}{2011}).
	
	\bibitem{Fang-2012}
	\bibinfo{author}{Fang, K.}, \bibinfo{author}{Yu, Z.} \& \bibinfo{author}{Fan,
		S.}
	\newblock \bibinfo{title}{Realizing effective magnetic field for photons by
		controlling the phase of dynamic modulation}.
	\newblock \emph{\bibinfo{journal}{Nature Photonics}}
	\textbf{\bibinfo{volume}{6}}, \bibinfo{pages}{782} (\bibinfo{year}{2012}).
	
	\bibitem{Rechtsman-2013}
	\bibinfo{author}{Rechtsman, M.~C.} \emph{et~al.}
	\newblock \bibinfo{title}{Photonic floquet topological insulators}.
	\newblock \emph{\bibinfo{journal}{Nature}} \textbf{\bibinfo{volume}{496}},
	\bibinfo{pages}{196} (\bibinfo{year}{2013}).
	
	\bibitem{Khanikaev2013}
	\bibinfo{author}{Khanikaev, A.~B.} \emph{et~al.}
	\newblock \bibinfo{title}{Photonic topological insulators}.
	\newblock \emph{\bibinfo{journal}{Nature Materials}}
	\textbf{\bibinfo{volume}{12}}, \bibinfo{pages}{233--239}
	(\bibinfo{year}{2013}).
	
	\bibitem{Lu-2014}
	\bibinfo{author}{Lu, L.}, \bibinfo{author}{Joannopoulos, J.~D.} \&
	\bibinfo{author}{Solja\v{c}i\'{c}, M.}
	\newblock \bibinfo{title}{Topological photonics}.
	\newblock \emph{\bibinfo{journal}{Nature Photonics}}
	\textbf{\bibinfo{volume}{8}}, \bibinfo{pages}{821} (\bibinfo{year}{2014}).
	
	\bibitem{Carusotto2013}
	\bibinfo{author}{Carusotto, I.} \& \bibinfo{author}{Ciuti, C.}
	\newblock \bibinfo{title}{Quantum fluids of light}.
	\newblock \emph{\bibinfo{journal}{Reviews of Modern Physics}}
	\textbf{\bibinfo{volume}{85}}, \bibinfo{pages}{299} (\bibinfo{year}{2013}).
	
	\bibitem{CedraMendez2010}
	\bibinfo{author}{Cerda-M\'endez, E.~A.} \emph{et~al.}
	\newblock \bibinfo{title}{Polariton condensation in dynamic acoustic lattices}.
	\newblock \emph{\bibinfo{journal}{Phys. Rev. Lett.}}
	\textbf{\bibinfo{volume}{105}}, \bibinfo{pages}{116402}
	(\bibinfo{year}{2010}).
	
	\bibitem{Kim2013}
	\bibinfo{author}{Kim, N.~Y.} \emph{et~al.}
	\newblock \bibinfo{title}{Exciton–polariton condensates near the dirac point
		in a triangular lattice}.
	\newblock \emph{\bibinfo{journal}{New Journal of Physics}}
	\textbf{\bibinfo{volume}{15}}, \bibinfo{pages}{035032}
	(\bibinfo{year}{2013}).
	
	\bibitem{Jacqmin2014}
	\bibinfo{author}{Jacqmin, T.} \emph{et~al.}
	\newblock \bibinfo{title}{Direct observation of dirac cones and a flatband in a
		honeycomb lattice for polaritons}.
	\newblock \emph{\bibinfo{journal}{Phys. Rev. Lett.}}
	\textbf{\bibinfo{volume}{112}}, \bibinfo{pages}{116402}
	(\bibinfo{year}{2014}).
	
	\bibitem{Baboux2016}
	\bibinfo{author}{Baboux, F.} \emph{et~al.}
	\newblock \bibinfo{title}{Bosonic condensation and disorder-induced
		localization in a flat band}.
	\newblock \emph{\bibinfo{journal}{Phys. Rev. Lett.}}
	\textbf{\bibinfo{volume}{116}}, \bibinfo{pages}{066402}
	(\bibinfo{year}{2016}).
	
	\bibitem{Liew2011}
	\bibinfo{author}{Liew, T.}, \bibinfo{author}{Shelykh, I.} \&
	\bibinfo{author}{Malpuech, G.}
	\newblock \bibinfo{title}{Polaritonic devices}.
	\newblock \emph{\bibinfo{journal}{Physica E}} \textbf{\bibinfo{volume}{43}},
	\bibinfo{pages}{1543} (\bibinfo{year}{2011}).
	
	\bibitem{Sanvitto2016}
	\bibinfo{author}{D.~Sanvitto, D.} \& \bibinfo{author}{K\'{e}na-Cohen, S.}
	\newblock \bibinfo{title}{The road towards polaritonic devices}.
	\newblock \emph{\bibinfo{journal}{Nature Materials}}
	\textbf{\bibinfo{volume}{15}}, \bibinfo{pages}{1061} (\bibinfo{year}{2016}).
	
	\bibitem{Karzig-PRX-2015}
	\bibinfo{author}{Karzig, T.}, \bibinfo{author}{Bardyn, C.-E.},
	\bibinfo{author}{Lindner, N.~H.} \& \bibinfo{author}{Refael, G.}
	\newblock \bibinfo{title}{Topological polaritons}.
	\newblock \emph{\bibinfo{journal}{Phys. Rev. X}} \textbf{\bibinfo{volume}{5}},
	\bibinfo{pages}{031001} (\bibinfo{year}{2015}).
	
	\bibitem{Bardyn-PRB-2015}
	\bibinfo{author}{Bardyn, C.-E.}, \bibinfo{author}{Karzig, T.},
	\bibinfo{author}{Refael, G.} \& \bibinfo{author}{Liew, T. C.~H.}
	\newblock \bibinfo{title}{Topological polaritons and excitons in garden-variety
		systems}.
	\newblock \emph{\bibinfo{journal}{Phys. Rev. B}} \textbf{\bibinfo{volume}{91}},
	\bibinfo{pages}{161413} (\bibinfo{year}{2015}).
	
	\bibitem{Nalitov-Z}
	\bibinfo{author}{Nalitov, A.~V.}, \bibinfo{author}{Solnyshkov, D.~D.} \&
	\bibinfo{author}{Malpuech, G.}
	\newblock \bibinfo{title}{Polariton $\mathbb{Z}$ topological insulator}.
	\newblock \emph{\bibinfo{journal}{Phys. Rev. Lett.}}
	\textbf{\bibinfo{volume}{114}}, \bibinfo{pages}{116401}
	(\bibinfo{year}{2015}).
	
	\bibitem{Yi-PRB-2016}
	\bibinfo{author}{Yi, K.} \& \bibinfo{author}{Karzig, T.}
	\newblock \bibinfo{title}{Topological polaritons from photonic dirac cones
		coupled to excitons in a magnetic field}.
	\newblock \emph{\bibinfo{journal}{Phys. Rev. B}} \textbf{\bibinfo{volume}{93}},
	\bibinfo{pages}{104303} (\bibinfo{year}{2016}).
	
	\bibitem{Bardyn-PRB-2016}
	\bibinfo{author}{Bardyn, C.-E.}, \bibinfo{author}{Karzig, T.},
	\bibinfo{author}{Refael, G.} \& \bibinfo{author}{Liew, T. C.~H.}
	\newblock \bibinfo{title}{Chiral bogoliubov excitations in nonlinear bosonic
		systems}.
	\newblock \emph{\bibinfo{journal}{Phys. Rev. B}} \textbf{\bibinfo{volume}{93}},
	\bibinfo{pages}{020502} (\bibinfo{year}{2016}).
	
	\bibitem{Janot-PRB-2016}
	\bibinfo{author}{Janot, A.}, \bibinfo{author}{Rosenow, B.} \&
	\bibinfo{author}{Refael, G.}
	\newblock \bibinfo{title}{Topological polaritons in a quantum spin hall
		cavity}.
	\newblock \emph{\bibinfo{journal}{Phys. Rev. B}} \textbf{\bibinfo{volume}{93}},
	\bibinfo{pages}{161111} (\bibinfo{year}{2016}).
	
	\bibitem{Gulevich-kagome}
	\bibinfo{author}{Gulevich, D.~R.}, \bibinfo{author}{Yudin, D.},
	\bibinfo{author}{Iorsh, I.~V.} \& \bibinfo{author}{Shelykh, I.~A.}
	\newblock \bibinfo{title}{Kagome lattice from an exciton-polariton
		perspective}.
	\newblock \emph{\bibinfo{journal}{Phys. Rev. B}} \textbf{\bibinfo{volume}{94}},
	\bibinfo{pages}{115437} (\bibinfo{year}{2016}).
	
	\bibitem{Lumer-PRL-2013}
	\bibinfo{author}{Lumer, Y.}, \bibinfo{author}{Plotnik, Y.},
	\bibinfo{author}{Rechtsman, M.~C.} \& \bibinfo{author}{Segev, M.}
	\newblock \bibinfo{title}{Self-localized states in photonic topological
		insulators}.
	\newblock \emph{\bibinfo{journal}{Phys. Rev. Lett.}}
	\textbf{\bibinfo{volume}{111}}, \bibinfo{pages}{243905}
	(\bibinfo{year}{2013}).
	
	\bibitem{Ostrovskaya2013}
	\bibinfo{author}{Ostrovskaya, E.~A.}, \bibinfo{author}{Abdullaev, J.},
	\bibinfo{author}{Fraser, M.~D.}, \bibinfo{author}{Desyatnikov, A.~S.} \&
	\bibinfo{author}{Kivshar, Y.~S.}
	\newblock \bibinfo{title}{Self-localization of polariton condensates in
		periodic potentials}.
	\newblock \emph{\bibinfo{journal}{Phys. Rev. Lett.}}
	\textbf{\bibinfo{volume}{110}}, \bibinfo{pages}{170407}
	(\bibinfo{year}{2013}).
	
	\bibitem{Hadad-PRB-2016}
	\bibinfo{author}{Hadad, Y.}, \bibinfo{author}{Khanikaev, A.~B.} \&
	\bibinfo{author}{Al\`u, A.}
	\newblock \bibinfo{title}{Self-induced topological transitions and edge states
		supported by nonlinear staggered potentials}.
	\newblock \emph{\bibinfo{journal}{Phys. Rev. B}} \textbf{\bibinfo{volume}{93}},
	\bibinfo{pages}{155112} (\bibinfo{year}{2016}).
	
	\bibitem{Bleu-PRB-2016}
	\bibinfo{author}{Bleu, O.}, \bibinfo{author}{Solnyshkov, D.~D.} \&
	\bibinfo{author}{Malpuech, G.}
	\newblock \bibinfo{title}{Interacting quantum fluid in a polariton chern
		insulator}.
	\newblock \emph{\bibinfo{journal}{Phys. Rev. B}} \textbf{\bibinfo{volume}{93}},
	\bibinfo{pages}{085438} (\bibinfo{year}{2016}).
	
	\bibitem{Kartashov-OL-2016}
	\bibinfo{author}{Kartashov, Y.~V.} \& \bibinfo{author}{Skryabin, D.~V.}
	\newblock \bibinfo{title}{Two-dimensional lattice solitons in polariton
		condensates with spin-orbit coupling}.
	\newblock \emph{\bibinfo{journal}{Opt. Lett.}} \textbf{\bibinfo{volume}{41}},
	\bibinfo{pages}{5043--5046} (\bibinfo{year}{2016}).
	
	\bibitem{DGulevich-Meissner}
	\bibinfo{author}{Gulevich, D.~R.}, \bibinfo{author}{Skryabin, D.~V.},
	\bibinfo{author}{Alodjants, A.~P.} \& \bibinfo{author}{Shelykh, I.~A.}
	\newblock \bibinfo{title}{Topological spin meissner effect in spinor
		exciton-polariton condensate: Constant amplitude solutions, half-vortices,
		and symmetry breaking}.
	\newblock \emph{\bibinfo{journal}{Phys. Rev. B}} \textbf{\bibinfo{volume}{94}},
	\bibinfo{pages}{115407} (\bibinfo{year}{2016}).
	
	\bibitem{Ablowitz-PRA-2014}
	\bibinfo{author}{Ablowitz, M.~J.}, \bibinfo{author}{Curtis, C.~W.} \&
	\bibinfo{author}{Ma, Y.-P.}
	\newblock \bibinfo{title}{Linear and nonlinear traveling edge waves in optical
		honeycomb lattices}.
	\newblock \emph{\bibinfo{journal}{Phys. Rev. A}} \textbf{\bibinfo{volume}{90}},
	\bibinfo{pages}{023813} (\bibinfo{year}{2014}).
	
	\bibitem{Leykam-PRL-2016}
	\bibinfo{author}{Leykam, D.} \& \bibinfo{author}{Chong, Y.~D.}
	\newblock \bibinfo{title}{Edge solitons in nonlinear-photonic topological
		insulators}.
	\newblock \emph{\bibinfo{journal}{Phys. Rev. Lett.}}
	\textbf{\bibinfo{volume}{117}}, \bibinfo{pages}{143901}
	(\bibinfo{year}{2016}).
	
	\bibitem{Kartashov-Optica-2016}
	\bibinfo{author}{Kartashov, Y.~V.} \& \bibinfo{author}{Skryabin, D.~V.}
	\newblock \bibinfo{title}{Modulational instability and solitary waves in
		polariton topological insulators}.
	\newblock \emph{\bibinfo{journal}{Optica 3}} \bibinfo{pages}{1228}
	(\bibinfo{year}{2016}).
	
	\bibitem{CedraMendez2016}
	\bibinfo{author}{Cerda-M\'endez, E.~A.} \emph{et~al.}
	\newblock \bibinfo{title}{Exciton-polariton gap solitons in two-dimensional
		lattices}.
	\newblock \emph{\bibinfo{journal}{Phys. Rev. Lett.}}
	\textbf{\bibinfo{volume}{111}}, \bibinfo{pages}{146401}
	(\bibinfo{year}{2013}).
	
	\bibitem{Soln-PRL-2017}
	\bibinfo{author}{Solnyshkov, D.~D.}, \bibinfo{author}{Bleu, O.},
	\bibinfo{author}{Teklu, B.} \& \bibinfo{author}{Malpuech, G.}
	\newblock \bibinfo{title}{Chirality of topological gap solitons in bosonic
		dimer chains}.
	\newblock \emph{\bibinfo{journal}{Phys. Rev. Lett.}}
	\textbf{\bibinfo{volume}{118}}, \bibinfo{pages}{023901}
	(\bibinfo{year}{2017}).
	
	\bibitem{Rajaraman}
	\bibinfo{author}{Rajaraman, R.}
	\newblock \emph{\bibinfo{title}{Solitons and Instantons: An Introduction to
			Solitons and Instantons in Quantum Field Theory}}
	(\bibinfo{publisher}{Elsevier Science}, \bibinfo{year}{1987}).
	
	\bibitem{Vasc-APL-2011}
	\bibinfo{author}{de~Vasconcellos, S.~M.} \emph{et~al.}
	\newblock \bibinfo{title}{Spatial, spectral, and polarization properties of
		coupled micropillar cavities}.
	\newblock \emph{\bibinfo{journal}{Appl. Phys. Lett.}}
	\textbf{\bibinfo{volume}{99}}, \bibinfo{pages}{101103}
	(\bibinfo{year}{2011}).
	
	\bibitem{Nalitov-PRL-2015}
	\bibinfo{author}{Nalitov, A.~V.}, \bibinfo{author}{Malpuech, G.},
	\bibinfo{author}{Ter\ifmmode~\mbox{\c{c}}\else \c{c}\fi{}as, H.} \&
	\bibinfo{author}{Solnyshkov, D.~D.}
	\newblock \bibinfo{title}{Spin-orbit coupling and the optical spin hall effect
		in photonic graphene}.
	\newblock \emph{\bibinfo{journal}{Phys. Rev. Lett.}}
	\textbf{\bibinfo{volume}{114}}, \bibinfo{pages}{026803}
	(\bibinfo{year}{2015}).
	
	\bibitem{Sala-PRX-2015}
	\bibinfo{author}{Sala, V.~G.} \emph{et~al.}
	\newblock \bibinfo{title}{Spin-orbit coupling for photons and polaritons in
		microstructures}.
	\newblock \emph{\bibinfo{journal}{Phys. Rev. X}} \textbf{\bibinfo{volume}{5}},
	\bibinfo{pages}{011034} (\bibinfo{year}{2015}).
	
	\bibitem{Hasan-RMP-2010}
	\bibinfo{author}{Hasan, M.~Z.} \& \bibinfo{author}{Kane, C.~L.}
	\newblock \bibinfo{title}{Colloquium: Topological insulators}.
	\newblock \emph{\bibinfo{journal}{Rev. Mod. Phys.}}
	\textbf{\bibinfo{volume}{82}}, \bibinfo{pages}{3045--3067}
	(\bibinfo{year}{2010}).
	
	\bibitem{Fruchart}
	\bibinfo{author}{Fruchart, M.} \& \bibinfo{author}{Carpentier, D.}
	\newblock \bibinfo{title}{An introduction to topological insulators}.
	\newblock \emph{\bibinfo{journal}{Comptes Rendus Physique}}
	\textbf{\bibinfo{volume}{14}}, \bibinfo{pages}{779 -- 815}
	(\bibinfo{year}{2013}).
	\newblock \bibinfo{note}{Topological insulators / Isolants
		topologiquesTopological insulators / Isolants topologiques}.
	
	\bibitem{Vladimirova}
	\bibinfo{author}{Vladimirova, M.} \emph{et~al.}
	\newblock \bibinfo{title}{Polarization controlled nonlinear transmission of
		light through semiconductor microcavities}.
	\newblock \emph{\bibinfo{journal}{Phys. Rev. B}} \textbf{\bibinfo{volume}{79}},
	\bibinfo{pages}{115325} (\bibinfo{year}{2009}).
	
	\bibitem{LL-III}
	\bibinfo{author}{Landau, L.~D.} \& \bibinfo{author}{Lifshitz, L.~M.}
	\newblock \emph{\bibinfo{title}{Quantum Mechanics, Third Edition:
			Non-Relativistic Theory (Volume 3)}} (\bibinfo{publisher}{Pregamon Press},
	\bibinfo{year}{1977}).
	
	\bibitem{Ablowitz}
	\bibinfo{author}{Ablowitz, M.~J.}
	\newblock \emph{\bibinfo{title}{Nonlinear Dispersive Waves: Asymptotic Analysis
			and Solitons}} (\bibinfo{publisher}{Cambridge University Press},
	\bibinfo{year}{2011}).
	
	\bibitem{Milicevic}
	\bibinfo{author}{Milićević, M.} \emph{et~al.}
	\newblock \bibinfo{title}{Edge states in polariton honeycomb lattices}.
	\newblock \emph{\bibinfo{journal}{2D Materials}} \textbf{\bibinfo{volume}{2}},
	\bibinfo{pages}{034012} (\bibinfo{year}{2015}).
	
	\bibitem{Galbiati}
	\bibinfo{author}{Galbiati, M.} \emph{et~al.}
	\newblock \bibinfo{title}{Polariton condensation in photonic molecules}.
	\newblock \emph{\bibinfo{journal}{Phys. Rev. Lett.}}
	\textbf{\bibinfo{volume}{108}}, \bibinfo{pages}{126403}
	(\bibinfo{year}{2012}).
	
	\bibitem{Abbarchi}
	\bibinfo{author}{Abbarchi, M.} \emph{et~al.}
	\newblock \bibinfo{title}{Macroscopic quantum self-trapping and josephson
		oscillations of exciton polaritons}.
	\newblock \emph{\bibinfo{journal}{Nature Physics}}
	\textbf{\bibinfo{volume}{9}}, \bibinfo{pages}{275–279}
	(\bibinfo{year}{2013}).
	
	\bibitem{Duff-APL}
	\bibinfo{author}{Dufferwiel, S.} \emph{et~al.}
	\newblock \bibinfo{title}{Tunable polaritonic molecules in an open microcavity
		system}.
	\newblock \emph{\bibinfo{journal}{Applied Physics Letters}}
	\textbf{\bibinfo{volume}{107}}, \bibinfo{pages}{201106}
	(\bibinfo{year}{2015}).
	
	\bibitem{Kuwata-PRL-1997}
	\bibinfo{author}{Kuwata-Gonokami, M.} \emph{et~al.}
	\newblock \bibinfo{title}{Parametric scattering of cavity polaritons}.
	\newblock \emph{\bibinfo{journal}{Phys. Rev. Lett.}}
	\textbf{\bibinfo{volume}{79}}, \bibinfo{pages}{1341--1344}
	(\bibinfo{year}{1997}).
	
	\bibitem{Savvidis-PRL-2000}
	\bibinfo{author}{Savvidis, P.~G.} \emph{et~al.}
	\newblock \bibinfo{title}{Angle-resonant stimulated polariton amplifier}.
	\newblock \emph{\bibinfo{journal}{Phys. Rev. Lett.}}
	\textbf{\bibinfo{volume}{84}}, \bibinfo{pages}{1547--1550}
	(\bibinfo{year}{2000}).
	
	\bibitem{Ciuti-PRB-2001}
	\bibinfo{author}{Ciuti, C.}, \bibinfo{author}{Schwendimann, P.} \&
	\bibinfo{author}{Quattropani, A.}
	\newblock \bibinfo{title}{Parametric luminescence of microcavity polaritons}.
	\newblock \emph{\bibinfo{journal}{Phys. Rev. B}} \textbf{\bibinfo{volume}{63}},
	\bibinfo{pages}{041303} (\bibinfo{year}{2001}).
	
	\bibitem{Dasbach-PRB-2001}
	\bibinfo{author}{Dasbach, G.}, \bibinfo{author}{Schwab, M.},
	\bibinfo{author}{Bayer, M.} \& \bibinfo{author}{Forchel, A.}
	\newblock \bibinfo{title}{Parametric polariton scattering in microresonators
		with three-dimensional optical confinement}.
	\newblock \emph{\bibinfo{journal}{Phys. Rev. B}} \textbf{\bibinfo{volume}{64}},
	\bibinfo{pages}{201309} (\bibinfo{year}{2001}).
	
	\bibitem{Tartakovskii-PRB-2002}
	\bibinfo{author}{Tartakovskii, A.~I.} \emph{et~al.}
	\newblock \bibinfo{title}{Polariton parametric scattering processes in
		semiconductor microcavities observed in continuous wave experiments}.
	\newblock \emph{\bibinfo{journal}{Phys. Rev. B}} \textbf{\bibinfo{volume}{65}},
	\bibinfo{pages}{081308} (\bibinfo{year}{2002}).
	
\end{thebibliography}

\section*{Acknowledgements}
D.Y. acknowledges support from RFBR project 16-32-60040. I.V.I. acknowledges support from RFBR Project 15-02-08949. I.A.S. and I.V.I acknowledge support from Rannis projects 141241-051 and 163082-051. D.V.S. acknowledges support from ITMO University  through the grant of the Government of the Russian Federation (074-U01). 

\section*{Author contributions statement}
D.Y. conceived the idea of the paper and contributed to the theory. 
D.R.G. wrote the main manuscript text, performed numerical analysis and prepared figures 1-2. 
D.V.S. contributed to the theory. I.V.I. and I.A.S. contributed to numerical calculations and introduction. All authors reviewed the manuscript.
 
\section*{Additional information}


\textbf{Competing financial interests}
The authors declare no competing financial interests.

\end{document}